\documentclass[draftcls,onecolumn,10pt,twoside]{IEEEtran}

\usepackage{ifpdf}

\usepackage{cite}

\usepackage{graphicx}

\ifCLASSINFOpdf
\else
\fi
\usepackage[cmex10]{amsmath}
\usepackage{amssymb,amsbsy}

\usepackage{algorithmic}

\usepackage{array}
\usepackage{dblfloatfix}
 \usepackage{url}

\begin{document}
%
\title{Adaptive Zero Reaction Motion Control for Free-Floating Space Manipulators}
%
%
%

\author{Shuanfeng~Xu,
        Hanlei~Wang,
        Duzhou~Zhang,
        and~Baohua~Yang                  
\thanks{S. Xu, H. Wang and D. Zhang are with the Science and Technology on Space Intelligent Control Laboratory,
   Beijing Institute of Control Engineering, Beijing 100190, China (e-mail: strivesfxu@gmail.com; hlwang.bice@gmail.com; zhangduzhou@gmail.com).
   }
\thanks{B. Yang is with China Aerospace Science and Technology Corporation, Beijing, 100048, China (e-mail: bhyang.cast@gmail.com).}
\thanks{This work was supported by the National Natural Science Foundation of China under Grants 61004058 and 61374060,
 and the National Key Basic Research Program (973) of China under Grant 2013CB733100.}}

\maketitle

\begin{abstract}
This paper investigates adaptive zero reaction motion control for free-floating space manipulators with uncertain kinematics and dynamics.
The challenge in deriving the adaptive reaction null-space (RNS) based control scheme is that it is difficult to obtain a linear expression, which is the basis of the adaptive control. The main contribution of this paper is that we skillfully obtain such a linear expression, based on which, an adaptive version of the RNS-based controller (referred to as the adaptive zero reaction motion controller in the sequel) is developed at the velocity level, taking into account both the kinematic and dynamic uncertainties. It is shown that the proposed controller achieves both the spacecraft attitude regulation and end-effector trajectory tracking.
The performance of the proposed adaptive controller is shown by numerical simulations with a planar 3-DOF (degree-of-freedom) space manipulator.
\end{abstract}


%
\IEEEpeerreviewmaketitle

\section{Introduction}

In many space tasks (e.g., capture and maintenance of a tumbling/failed spacecraft), it would be dangerous if relying on extra-vehicular activities (EVA) performed by astronauts.
A much safer strategy is to use robot manipulators, and in fact they are now playing a more and more important role in space exploration (see, e.g., \cite{Stoll:2009, Imaida:2001, Hirzinger:2004, Weismuller:2006}).
The base of the manipulator (i.e., the spacecraft) in space is usually not fixed, and the system consisting of the manipulator and the spacecraft is referred to as the space manipulator in the literature.
When manipulating various objects (e.g., a noncooperative target), it is inevitable to encounter parametric uncertainties, which has the tendency of lowering the tracking accuracy of the system \cite{Slotine:1989}. Adaptive control, as a standard control methodology, is a qualified approach to handle parametric uncertainties \cite{Slotine:1991}.

Among the control modes of space manipulators, free-floating space manipulators (FFSM) have their potential advantages, e.g., non-renewable fuel on the spacecraft can be saved and the safety of close-range manipulation can be ensured \cite{Papadopoulos:1990}.
It is known that in a free-floating space manipulator, the motion of the spacecraft will evolve under the dynamic reaction due to that of the manipulator, and the evolution of the whole system is governed by the principle of momentum conservation.
For the end-effector tracking problem without consideration of the spacecraft attitude, many adaptive control algorithms have been proposed (e.g., \cite{Gu:1995}, \cite{Wang:2012}). Specifically, the tracking objective  is realized by the prediction error based approach in \cite{Wang:2012} with the uncertainties of both the kinematics and dynamics being taken into consideration.
However,
in practice the spacecraft attitude maintenance is a major concern since the communication with the Earth can be carried out only when the spacecraft antenna points to the Earth (guaranteed by the attitude maintenance control) \cite{Aghili:2009}.
Hence, joint motion algorithms for space manipulators without reaction to the spacecraft are highly preferred. On the other hand, the manipulator end-effector is usually required to track some trajectory in Cartesian space when executing On-orbit servicing (OOS). Thus, it is meaningful to realize coordinated spacecraft/manipulator motion control.

Many researchers have studied coordination control of a manipulator and its free-floating base.
Vafa and Dubowsky proposed joint cyclic motion algorithm so that the spacecraft orientation is maintained \cite{Vafa:1990}. Nakamura and Mukherjee presented an algorithm
to achieve the regulation of both the spacecraft attitude and the manipulator joint angles simultaneously, where the stability of the system is analyzed by the Lyapunov method \cite{Nakamura:1991}. The motion planning for a system of coupled rigid bodies is investigated in \cite{Fernandes:1994}, which is claimed to be applicable in space robotic applications  \cite{Fernandes:1994}.
Dubowsky and Torres proposed a joint motion scheme using EDM (enhanced disturbance map), which ensures that the disturbance on the spacecraft attitude is minimized \cite{Dubowsky:1991}.
Yamada developed a closed joint trajectory for the manipulator relying on the variational approach, where the spacecraft attitude can be regulated to any desired value \cite{Yamada:1993}.
Suzuki and Nakamura devised ``spiral motion'' for the end-effector such that the spacecraft attitude and the manipulator joint position are regulated to their desired values (constant), which is unfortunately an approximate method \cite{Suzuki:1996}.
The point-to-point planning method provided in \cite{Tortopidis:2007} employs the smooth high-order polynomials to achieve the regulation of both the spacecraft attitude and manipulator joint positions, without requiring the aforementioned cyclic robot motion, and in order to ensure the existence of this kind of joint trajectory, the desired spacecraft attitude must lie in certain bounded region.
The reaction null-space algorithm proposed in \cite{Nenchev:1992}, unlike the results mentioned above, achieves both the end-effector trajectory tracking and the attitude keeping with the use of the manipulator DOFs only.
As is now well recognized, the redundancy of the manipulator is a prerequisite in realizing the attitude regulation using the RNS algorithm, whose possible advantages, in comparison with the other approaches, may be that it no longer needs the cyclic manipulator motion and that it imposes no constraint on the desired attitude variation.

The concept of reaction null-space dates back to \cite{Nenchev:1992}. Then, the RNS control law was used to suppress the vibrating motion associated with a system composed of a flexible structure and a manipulator mounted on it \cite{Nenchev:1999}.
A kinematic control scheme based on reaction null-space can achieve zero reaction manipulation, referred to as zero reaction maneuver (ZRM), where a combined inertia and Jacobian matrix is introduced \cite{Yoshida:2001}. The RNS-based zero reaction manipulation was carried out and verified in the ETS-VII project \cite{Yoshida:2003}.
Later, the RNS-based controller was used for JEMRMS  end-effector trajectory tracking with zero reaction motion and vibration suppression in \cite{Fukazu:2009}.
A zero reaction trajectory generation strategy was developed without affecting the spacecraft attitude for the capture of a target by a 2-DOF manipulator \cite{Piersigilli:2010}. However, it should be noted that the methods proposed above require the exact knowledge of both the system kinematics and dynamics.

In the presence of parameter uncertainties or variations,
the challenge is that it is difficult to find an appropriate linear expression with respect to the uncertain parameters,  which is the basis of designing parameter adaptation law.
To the best of our knowledge, the only attempt to resolve this problem occurs in \cite{Nguyen_Huynh:2013}, where an adaptive zero reaction motion algorithm for space manipulators was proposed with the dynamic uncertainties being taken into consideration.
However, the work in \cite{Nguyen_Huynh:2013} has not proved why the control objective can be achieved if the actual joint velocity of the manipulator is identical to the designed velocity, and in addition, the designed velocity includes an undesirable algebraic loop.
Furthermore, in the work of \cite{Nguyen_Huynh:2013}, only the regulation of the spacecraft angular velocity is considered, in which case, the spacecraft attitude will possibly deviate from its desired value during the adaptive control.

In this study, we skillfully obtain a linear expression with respect to the uncertain parameters.
Based on this expression, we propose an adaptive zero reaction motion control algorithm that can deal with both the dynamic and kinematic uncertainties.
These uncertainties could arise from the lack of accurate knowledge of the parameters of the manipulator or the unknown target that is captured by the manipulator.
In contrast to the work of \cite{Nguyen_Huynh:2013}, the proposed algorithm can regulate the attitude of the spacecraft during the adaptive control.
Step by step, two joint motion control algorithms are designed at velocity level to respectively achieve 1) the spacecraft attitude regulation with simultaneous optimization of a rather general performance index, and 2) both the spacecraft attitude regulation and end-effector trajectory tracking.
In summary, the main contribution of our work is that we give an adaptive zero reaction motion controller for FFSMs, extending the existing results (e.g., \cite{Nenchev:1992}, \cite{Yoshida:2001}) to the unknown parameter case.
The handling of the kinematic uncertainties can be considered to be an extension of the results for fixed-base robots in \cite{Cheah:2006a,Cheah:2006b} and the one for FFSMs in \cite{Wang:2012}.
We also take into consideration the case of the presence of the nonzero initial linear and angular momenta in the system.
A preliminary version of the paper appears in \cite{Xu:2013},
which only considers the case of zero initial momenta, and here, we extend this preliminary result to cover the case where there are nonzero initial momenta.

The rest of the paper is organized as follows. In Section II, the dynamics and kinematics which characterize a free-floating space manipulator are given, and the derivation of RNS are presented.
Then, the adaptive zero reaction motion controller is developed in Section III.
To demonstrate the effectiveness of the proposed method, simulation results are shown in Section IV.
Finally, the conclusions and future work are stated in Section V.

\section{PRELIMINARIES}

\subsection{Dynamics of FFSMs}

The equations of motion of a free-floating space robot explicitly including the rotational motion of the spacecraft are described by \cite{Xu:1992,Xu:1994}
\begin{equation} \label{1}
{\begin{bmatrix} \mathbf{H}_{b} & \mathbf{H}_{bm} \\[4pt]
                 \mathbf{H}^{\mathbf{T}}_{bm} & \mathbf{H}_{m} \end{bmatrix}}
{\begin{bmatrix} \dot{\boldsymbol\omega}_{b} \\[4pt] \ddot{\boldsymbol\phi} \end{bmatrix}}
+ {\begin{bmatrix} \mathbf{c}_{b} \\[4pt] \mathbf{c}_{m} \end{bmatrix}}
= {\begin{bmatrix} \mathbf{0}_3 \\[4pt] \boldsymbol\tau_{m} \end{bmatrix}}
\end{equation}
where $\boldsymbol\omega_{b} \in {\mathbb{R}}^3$ denotes the angular velocity of the spacecraft with respect to the inertial frame expressed in the spacecraft frame,
$\boldsymbol\phi = \begin{bmatrix} \phi_{1}, \dots, \phi_{n} \end{bmatrix}^{\mathbf{T}}$ is the joint angle,
$\dot{\boldsymbol\phi}$ denotes the joint velocity,
$\mathbf{H}_{b} \in {\mathbb{R}^{3 \times 3}}$ is the inertia matrix of the spacecraft, $\mathbf{H}_{m} \in {\mathbb{R}}^{n \times n}$ is the inertia matrix of the manipulator, $\mathbf{H}_{bm} \in {\mathbb{R}}^{3 \times n}$ is the coupled inertia matrix between the spacecraft and the manipulator,
$\mathbf{c}_{b} \in {\mathbb{R}}^{3}$ and $\mathbf{c}_{m} \in {\mathbb{R}}^{n}$ are the Coriolis and centrifugal forces, and $\boldsymbol\tau_{m} \in {\mathbb{R}}^{n}$ is the manipulator joint torque.

In the case that there is nonzero initial angular momentum, the integral of the upper part of~(\ref{1}) with respect to time yields \cite{Yoshida:2003}
\begin{equation} \label{2}
\mathbf{R}_{b} ( \mathbf{H}_{b} \boldsymbol\omega_{b} + \mathbf{H}_{bm} \dot{\boldsymbol\phi} )
\doteq \bar{\mathbf{H}}_{b} \boldsymbol\omega_{b} + \bar{\mathbf{H}}_{bm} \dot{\boldsymbol\phi} = \mathbf{p}_0
\end{equation}
where $\mathbf{R}_{b} \in \mathrm{SO}(3)$ is the spacecraft orientation matrix with respect to the inertial frame,
$\bar{\mathbf{H}}_{b} = \mathbf{R}_{b} \mathbf{H}_{b}$, $\bar{\mathbf{H}}_{bm} = \mathbf{R}_{b} \mathbf{H}_{bm}$,
and $\mathbf{p}_0$ is the initial angular momentum of the space manipulator system.
The momentum conservation equation~(\ref{2}) is simpler than the equation of motion at acceleration level, yet reflects almost all aspects of the system dynamics \cite{Yoshida:1994}.

Equation~(\ref{2}) depends linearly on a dynamic parameter vector $\mathbf{a}_{d} = \begin{bmatrix} a_{d1},a_{d2},\dots,a_{di} \end{bmatrix}^{\mathbf{T}}$
and the initial angular momentum $\mathbf{p}_0$ \cite{Abiko:2007}
\begin{equation} \label{3}
\begin{aligned}
& \bar{\mathbf{H}}_{b} \boldsymbol\omega_{b} + \bar{\mathbf{H}}_{bm} \dot{\boldsymbol\phi} - \mathbf{p}_0   \\
& = { \begin{bmatrix} \mathbf{Y}_{d} (\boldsymbol\epsilon_b, \boldsymbol\phi, \boldsymbol\omega_{b}, \dot{\boldsymbol\phi}) ~~~ -\mathbf{E}_{3 \times 3} \end{bmatrix} }
  { \begin{bmatrix}  \mathbf{a}_{d} \\[4pt]
                     \mathbf{p}_0   \end{bmatrix} }   \\
& \doteq \bar{\mathbf{Y}}_{d} \bar{\mathbf{a}}_{d}
\end{aligned}
\end{equation}
where $\bar{\mathbf{Y}}_{d} = { \begin{bmatrix} \mathbf{Y}_{d} (\boldsymbol\epsilon_b, \boldsymbol\phi, \boldsymbol\omega_b, \dot{\boldsymbol\phi}) ~~~ -\mathbf{E}_{3 \times 3} \end{bmatrix} }$ is referred to as the generalized dynamic regressor matrix,
$\bar{\mathbf{a}}_{d} = { \begin{bmatrix}  \mathbf{a}^{\mathbf{T}}_{d} ~~~ \mathbf{p}^{\mathbf{T}}_0   \end{bmatrix} }^{\mathbf{T}}$ referred to as the generalized dynamic parameters,
$\boldsymbol\epsilon_b \in {\mathbb{R}}^{4}$ are quaternions used to represent the spacecraft attitude,
$\mathbf{E}_{3 \times 3}$ is the $3 \times 3$ identity matrix,
and $\mathbf{Y}_{d}(\boldsymbol\epsilon_b, \boldsymbol\phi, \boldsymbol\omega_b, \dot{\boldsymbol\phi}) \in {\mathbb{R}}^{3 \times i}$ is the regressor matrix when $\mathbf{p}_0 = \mathbf{0}$.

\subsection{Kinematics of FFSMs}

Denote by $m$ the dimension of the task space. The FFSM end-effector velocity $\dot{\mathbf{x}} \in {\mathbb{R}^{m}}$ in the inertial frame can be expressed as \cite{Umetani:1989}
\begin{equation} \label{4}
\dot{\mathbf{x}} = \mathbf{J}_{b} \boldsymbol{\omega}_{b} + \mathbf{J}_{m} {\dot{\boldsymbol\phi}} + \mathbf{v}_0
\end{equation}
where $\mathbf{J}_{b} \in {\mathbb{R}}^{m \times 3}$
and $\mathbf{J}_{m} \in {\mathbb{R}}^{m \times n}$
are the Jacobian matrices,
and the appearance of the constant initial translational motion term $\mathbf{v}_0 \in {\mathbb{R}^{m}}$ is due to the nonzero linear momentum.

The kinematic equation~(\ref{4}) depends linearly on a kinematic parameter vector
$\mathbf{a}_{k} = \begin{bmatrix} a_{k1},a_{k2},\dots,a_{kj} \end{bmatrix}^{\mathbf{T}}$ and $\mathbf{v}_0$ \cite{Ma:1995, Cheah:2006b}
\begin{equation} \label{5}
\begin{aligned}
\dot{\mathbf{x}} &= \mathbf{J}_{b} \boldsymbol{\omega}_{b} + \mathbf{J}_{m} {\dot{\boldsymbol\phi}} + \mathbf{v}_0  \\
&= {\begin{bmatrix} \mathbf{Y}_{k}(\boldsymbol\epsilon_b, \boldsymbol\phi, \boldsymbol\omega_b, \dot{\boldsymbol\phi}) ~~~\mathbf{E}_{m \times m} \end{bmatrix}}
      { \begin{bmatrix} \mathbf{a}_{k} \\[4pt]
                        \mathbf{v}_0      \end{bmatrix} }  \\
&  \doteq  \bar{\mathbf{Y}}_{k} \bar{\mathbf{a}}_{k}
\end{aligned}
\end{equation}
where $\bar{\mathbf{Y}}_{k} = { \begin{bmatrix} \mathbf{Y}_{k}(\boldsymbol\epsilon_b, \boldsymbol\phi, \boldsymbol\omega_b, \dot{\boldsymbol\phi}) ~~~ \mathbf{E}_{m \times m} \end{bmatrix} }$ is referred to as the generalized kinematic regressor matrix,
$\bar{\mathbf{a}}_{k} = { \begin{bmatrix} \mathbf{a}^{\mathbf{T}}_{k} ~~~ \mathbf{v}^{\mathbf{T}}_0      \end{bmatrix} }^\mathbf{T}$ referred to as the generalized kinematic parameters,
and $\mathbf{Y}_{k}(\boldsymbol\epsilon_b, \boldsymbol\phi, \boldsymbol\omega_b, \dot{\boldsymbol\phi}) \in {\mathbb{R}}^{m \times j}$ is the kinematic regressor matrix.

\subsection{Reaction null-space}

Following the work of \cite{Yoshida:2003}, we briefly describe the basic idea of the reaction null-space.

Assume that the initial angular momentum is zero, and letting $\boldsymbol\omega_b=\mathbf{0}$, we obtain from the angular momentum conservation equation~(\ref{2}) that
\begin{equation} \label{6}
\bar{\mathbf{H}}_{bm} \dot{\boldsymbol{\phi}} = \mathbf{0}.
\end{equation}

Equation~(\ref{6}) leads to the following solution
\begin{equation} \label{7}
\dot{\boldsymbol\phi}_r = (\mathbf{E}_{n \times n} - \bar{\mathbf{H}}^{+}_{bm} \bar{\mathbf{H}}_{bm}) \boldsymbol\zeta
\end{equation}
where $(\cdot)^{+} = (\cdot)^{\mathbf{T}} \left[(\cdot)(\cdot)^{\mathbf{T}}\right]^{-1}$ denotes the standard right pseudoinverse of $(\cdot)$,
and $\mathbf{E}_{n \times n}$ is the $n \times n$ identity matrix.
The vector $\boldsymbol\zeta$ is arbitrary and the null-space of the inertia matrix $\bar{\mathbf{H}}_{bm}$ is called the reaction null-space.
The matrix $\mathbf{T} = \mathbf{E}_{n \times n} - \bar{\mathbf{H}}^{+}_{bm} \bar{\mathbf{H}}_{bm}$ in~(\ref{7}) denotes the projector onto the null-space of the coupled inertia matrix $\bar{\mathbf{H}}_{bm}$.
The joint motion given by~(\ref{7}) can ensure zero disturbance to the spacecraft attitude.

\emph{REMARK} 1. Eq.~(\ref{7}) can not be linearly parameterized with respect to a group of physical parameters due to the advent of $ \bar{\mathbf{H}}^{+}_{bm}$, which is a great challenge for the application of the conventional adaptive control.

In this paper, we assume that there exists the reaction null-space.

\section{ADAPTIVE ZERO REACTION CONTROL}

In this section, we derive an adaptive zero reaction kinematic controller for FFSMs with uncertain kinematics and dynamics.

\subsection{Problem Formulation}

Assuming that there exists a fast enough dynamic control law,
seek an adaptive kinematic control law $\dot{\boldsymbol\phi}^{\ast}_r$
to achieve both the attitude regulation of the spacecraft and trajectory tracking of the manipulator end-effector.
That is, $\boldsymbol{\omega}_{b} \to \mathbf{0}$, $\mathbf{R}_{b} \to \mathbf{R}_{bd}$, $\Delta \mathbf{x} \to \mathbf{0}$ and $\Delta\dot{\mathbf{x}} \to \mathbf{0}$ as $t \to \infty$.

Here, $\Delta \mathbf{x} = \mathbf{x} - \mathbf{x}_{d}$ is the tracking error of the end-effector, and $\mathbf{x}_{d} \in \mathbb{R}^{m}$ is the desired trajectory of the end-effector.
The boundedness of $\mathbf{x}_{d}$, $\dot{\mathbf{x}}_{d}$, and $\ddot{\mathbf{x}}_{d}$ is assumed. $\mathbf{R}_{b}$ and $\mathbf{R}_{bd}$ are the current and desired attitude matrices of the spacecraft, respectively,
where the desired attitude matrix $\mathbf{R}_{bd}$ is constant.

In this work, we do not explicitly design the torque control input but assume the existence of a dynamic controller or joint velocity servo controller.
We assume that the joint velocity servo control is fast enough so that the actual manipulator joint velocity $\dot{\boldsymbol\phi}$ can track the designed joint velocity $\dot{\boldsymbol\phi}_r^\ast$ immediately, which means $\dot{\boldsymbol\phi} \equiv \dot{\boldsymbol\phi}_r^\ast$.

\subsection{Adaptive Controller Design Considering the Spacecraft Attitude Regulation}

In order to achieve the spacecraft attitude regulation for FFSM with uncertain dynamics and nonzero initial momenta, we propose the following kinematic control law
\begin{equation} \label{10}
\dot{\boldsymbol\phi}^{\ast}_r = (\mathbf{E}_{n \times n} - \hat{\bar{\mathbf{H}}}^{+}_{bm} \hat{\bar{\mathbf{H}}}_{bm}) \boldsymbol\zeta
+ \hat{\bar{\mathbf{H}}}^{+}_{bm} (\hat{\mathbf{p}}_0 + \hat{\bar{\mathbf{H}}}_{b} \lambda_{b} \Delta\boldsymbol{\epsilon}_{bv})
\end{equation}
where $\hat{\bar{\mathbf{H}}}_{bm}$ is obtained by replacing the dynamic parameters in $\bar{\mathbf{H}}_{bm}$ with their estimates, $\hat{\mathbf{p}}_0$ is the estimate of the initial angular momentum,
$\lambda_{b} > 0$ is a constant, and $\Delta\boldsymbol{\epsilon}_{bv}$ is the vector part of the error quaternion corresponding to the error attitude matrix $\Delta\mathbf{R}_{b} = \mathbf{R}^{\mathbf{T}}_{bd} \mathbf{R}_{b}$ \cite{Egeland:1994}.

Premultiplying both sides of~(\ref{10}) by $\hat{\bar{\mathbf{H}}}_{bm}$, we have
\begin{equation} \label{11}
\hat{\bar{\mathbf{H}}}_{bm} \dot{\boldsymbol\phi}^{\ast}_r = \hat{\mathbf{p}}_0 + \hat{\bar{\mathbf{H}}}_{b} \lambda_{b} \Delta\boldsymbol{\epsilon}_{bv}.
\end{equation}

Combining~(\ref{2}) and~(\ref{11}), we get
\begin{equation} \label{12}
-\hat{\bar{\mathbf{H}}}_{bm} \dot{\boldsymbol\phi}^{\ast}_r + \hat{\bar{\mathbf{H}}}_{b} \lambda_{b} \Delta\boldsymbol{\epsilon}_{bv}
= -\bar{\mathbf{H}}_{b} \boldsymbol\omega_{b} - \bar{\mathbf{H}}_{bm} \dot{\boldsymbol\phi} + \mathbf{p}_0 - \hat{\mathbf{p}}_0.
\end{equation}

Adding $\hat{\bar{\mathbf{H}}}_{b} \boldsymbol\omega_{b} + \hat{\bar{\mathbf{H}}}_{bm} \dot{\boldsymbol\phi}$ to both sides of~(\ref{12}), we have
\begin{equation} \label{13}
\begin{array}{ll}
& \hat{\bar{\mathbf{H}}}_{b} (\boldsymbol\omega_{b} + \lambda_{b} \Delta\boldsymbol{\epsilon}_{bv}) + \hat{\bar{\mathbf{H}}}_{bm} (\dot{\boldsymbol{\phi}} - \dot{\boldsymbol\phi}^{\ast}_r) \\[6pt]
& = \Delta \bar{\mathbf{H}}_{b} \boldsymbol\omega_{b} + \Delta \bar{\mathbf{H}}_{bm} \dot{\boldsymbol\phi} - \Delta \mathbf{p}_0 \\[6pt]
& = { \begin{bmatrix} \mathbf{Y}_{d} (\boldsymbol\epsilon_b, \boldsymbol\phi, \boldsymbol\omega_{b}, \dot{\boldsymbol\phi}) ~~~ -\mathbf{E}_{3 \times 3} \end{bmatrix} }
  { \begin{bmatrix}  \Delta \mathbf{a}_{d} \\[4pt]
                     \Delta \mathbf{p}_0   \end{bmatrix} }  \\[6pt]
& = \bar{\mathbf{Y}}_{d} \Delta \bar{\mathbf{a}}_{d}
\end{array}
\end{equation}
where $\Delta \bar{\mathbf{H}}_{b} = \hat{\bar{\mathbf{H}}}_{b} - \bar{\mathbf{H}}_{b}$, $\Delta \bar{\mathbf{H}}_{bm} = \hat{\bar{\mathbf{H}}}_{bm} - \bar{\mathbf{H}}_{bm}$, $\Delta \mathbf{p}_0 = \hat{\mathbf{p}}_0 - \mathbf{p}_0$,
and $\Delta \bar{\mathbf{a}}_{d} = \hat{\bar{\mathbf{a}}}_{d} - \bar{\mathbf{a}}_{d}$ is the generalized dynamic parameter estimation error.
Let
\begin{equation} \label{14}
\mathbf{y}_{1} = \hat{\bar{\mathbf{H}}}_{b} (\boldsymbol\omega_{b} + \lambda_{b} \Delta\boldsymbol{\epsilon}_{bv}) + \hat{\bar{\mathbf{H}}}_{bm} (\dot{\boldsymbol{\phi}} - \dot{\boldsymbol\phi}^{\ast}_r).
\end{equation}
We assume that the quaternions corresponding to the spacecraft attitude $\boldsymbol\epsilon_{b}$, the angular velocity of the spacecraft $\boldsymbol\omega_{b}$, the joint angle of the manipulator $\boldsymbol\phi$, and the joint velocity of the manipulator $\dot{\boldsymbol\phi}$ can be obtained from the sensors. Therefore, the signal $\mathbf{y}_{1}$ is measurable. For the attitude regulation problem, the desired value of $\boldsymbol\omega_{b}$ is zero,
and thus the regulation error of the angular velocity of the spacecraft is $\Delta\boldsymbol\omega_b = \boldsymbol\omega_{b} - \mathbf{0} = \boldsymbol\omega_{b}$, which means that $\mathbf{y}_{1}$ can be rewritten as
\begin{equation} \label{15}
\mathbf{y}_{1} = \hat{\bar{\mathbf{H}}}_{b} (\Delta\boldsymbol\omega_b + \lambda_{b} \Delta\boldsymbol\epsilon_{bv}) + \hat{\bar{\mathbf{H}}}_{bm} (\dot{\boldsymbol{\phi}} - \dot{\boldsymbol\phi}^{\ast}_r).
\end{equation}

Now the gradient estimator of the standard form is adopted to update the generalized dynamic parameter estimate $\hat{\bar{\mathbf{a}}}_{d}$, and the updating law is given by
\begin{equation} \label{16}
\dot{\hat{\bar{\mathbf{a}}}}_{d} = - \boldsymbol\Gamma_{d} \bar{\mathbf{Y}}^{\mathbf{T}}_{d} \mathbf{y}_{1}
\end{equation}
where $\boldsymbol\Gamma_{d}$ is a constant symmetric positive definite matrix. Based on the work of \cite{Slotine:1991}, we know that $\mathbf{y}_{1} \in \mathcal{L}_2$, and $\hat{\bar{\mathbf{a}}}_{d} \in \mathcal{L}_{\infty}$.

Differentiating~(\ref{14}) with respect to time, we get
\begin{equation} \label{21}
\begin{aligned}
\dot{\mathbf{y}}_{1} = & \dot{\hat{\bar{\mathbf{H}}}}_{b} (\boldsymbol\omega_{b} + \lambda_{b} \Delta\boldsymbol{\epsilon}_{bv})
+ \hat{\bar{\mathbf{H}}}_{b} (\dot{\boldsymbol\omega}_{b} + \lambda_{b} \Delta\dot{\boldsymbol{\epsilon}}_{bv}) \\[6pt]
& + \dot{\hat{\bar{\mathbf{H}}}}_{bm} (\dot{\boldsymbol{\phi}} - \dot{\boldsymbol\phi}^{\ast}_r)
+ \hat{\bar{\mathbf{H}}}_{bm} (\ddot{\boldsymbol{\phi}} - \ddot{\boldsymbol\phi}^{\ast}_r)
\end{aligned}
\end{equation}
where $\ddot{\boldsymbol\phi}^{\ast}_r$ is the time derivative of $\dot{\boldsymbol\phi}^{\ast}_r$.

Here, we introduce a sliding variable \cite{Egeland:1994}
\begin{equation} \label{22}
\mathbf{s}_{b} = \Delta\boldsymbol\omega_b + \lambda_{b} \Delta\boldsymbol{\epsilon}_{bv}.
\end{equation}

It is reasonable to assume that $\boldsymbol\zeta \in \mathcal{L}_{\infty}$, and then the boundedness of $\hat{\bar{\mathbf{a}}}_{d}$ gives that $\dot{\boldsymbol\phi}^{\ast}_r \in \mathcal{L}_{\infty}$ if $\hat{\bar{\mathbf{H}}}_{bm}$ has full row rank.
By assumption, we have that $\dot{\boldsymbol{\phi}} \equiv \dot{\boldsymbol\phi}^{\ast}_r$, which yields the conclusion that  $\dot{\boldsymbol{\phi}} \in \mathcal{L}_{\infty}$.
From~(\ref{2}), we obtain that $\boldsymbol\omega_{b} \in \mathcal{L}_{\infty}$, which leads to the result that $\Delta\dot{\boldsymbol{\epsilon}}_{bv} \in \mathcal{L}_{\infty}$.
From~(\ref{15}), we have that $\mathbf{s}_{b} \in \mathcal{L}_2$ if $\hat{\mathbf{H}}_{b}$ is uniformly positive definite.
Thus, based on the work of \cite{Egeland:1994}, we have that $\boldsymbol\omega_b \in \mathcal{L}_2$.
It is easy to know from~(\ref{14}) that $\mathbf{y}_{1} \in \mathcal{L}_{\infty}$, which yields that $\dot{\hat{\bar{\mathbf{a}}}}_{d} \in \mathcal{L}_{\infty}$.
The vector $\boldsymbol\zeta$ is usually a function of the spacecraft attitude $\boldsymbol\epsilon_b$ and the manipulator joint position $\boldsymbol\phi$ due to the property of the kinematic controller~(\ref{10}), which implies that $\dot{\boldsymbol\zeta} \in \mathcal{L}_{\infty}$.
Therefore, if $\hat{\bar{\mathbf{H}}}_{bm}$ has full row rank, we obtain that $\ddot{\boldsymbol\phi}^{\ast}_r \in \mathcal{L}_{\infty}$,
which implies that $\ddot{\boldsymbol{\phi}} \in \mathcal{L}_{\infty}$ from the assumption that $\dot{\boldsymbol\phi} \equiv \dot{\boldsymbol\phi}_r^\ast$.
From the upper part of~(\ref{1}), we know that $\dot{\boldsymbol\omega}_{b} \in \mathcal{L}_{\infty}$.
We obtain that $\dot{\mathbf{y}}_{1} \in \mathcal{L}_{\infty}$ from~(\ref{21}),
and thus the signal $\mathbf{y}_{1}$ is uniformly continuous.

So far, we have known that $\mathbf{y}_{1} \in \mathcal{L}_2$ and $\mathbf{y}_{1}$ is uniformly continuous, and from the properties of square-integrable and uniformly continuous functions \cite[p.232]{Desoer:1975}, we have that $\mathbf{y}_{1} \to \mathbf{0}$ as $t \to \infty$.
Then, we obtain from (\ref{15}) that $\mathbf{s}_{b} \to \mathbf{0}$ as $t \to \infty$
if $\hat{\mathbf{H}}_{b}$ is uniformly positive definite.
According to the analysis in the work of \cite{Egeland:1994},
the fact that $\mathbf{s}_{b} \to \mathbf{0}$ as $t \to \infty$ implies that
$\Delta\boldsymbol{\epsilon}_{bv} \to \mathbf{0}$ and $\Delta\boldsymbol\omega_b \to \mathbf{0}$ as $t \to \infty$,
which means that
$\boldsymbol\omega_{b} \to \mathbf{0}$ and $\mathbf{R}_{b} \to \mathbf{R}_{bd}$ as $t \to \infty$.

\emph{REMARK} 2. Here we assume that the estimate of the inertia matrix of the spacecraft $\hat{\mathbf{H}}_{b}$ is positive definite and that the estimate of the coupled inertia matrix $\hat{\bar{\mathbf{H}}}_{bm}$ has full row rank, which can possibly be guaranteed by the parameter projection algorithm \cite{Ioanou:1996, Wang:2011}.

We summarize the above analysis as the following theorem.
\newtheorem{theorem}{Theorem}
\begin{theorem}
The kinematic control law~(\ref{10}) and the parameter adaptation law~(\ref{16}) achieve the spacecraft attitude regulation provided that there exists a fast enough dynamic control law so that $\dot{\boldsymbol{\phi}} \equiv \dot{\boldsymbol\phi}^{\ast}_r$.
That is, $\boldsymbol{\omega}_{b} \to \mathbf{0}$ and $\mathbf{R}_{b} \to \mathbf{R}_{bd}$ as $t \to \infty$.
\end{theorem}

\subsection{Adaptive Controller Design Considering Both the Spacecraft Attitude Regulation and End-effector Trajectory Tracking}

When the FFSM is executing OOS, it is usually not enough to control only the spacecraft attitude. Under this circumstance, the end-effector of the FFSM is usually required to track a desired trajectory $\mathbf{x}_{d} \in \mathbb{R}^{m}$. Here, we assume that the space manipulator motions in a workspace where the dynamic singularity does not occur.

Let $d_{1}$ and $d_{2}(=m)$  be the number of task variables for the spacecraft task and the end-effector task, respectively. As long as the number of the manipulator joints $n$ is not smaller than the total number of task variables $d_{1}+d_{2}$, i.e., $n \ge d_{1}+d_{2}$, both the spacecraft attitude regulation and end-effector trajectory tracking can be achieved by making appropriate choice of $\boldsymbol{\zeta}$ \cite{Nenchev:1992}. So $n \ge d_{1}+d_{2}$ is assumed in this paper.

Next, we will exploit the property of $\boldsymbol{\zeta}$ to achieve both the spacecraft attitude regulation and end-effector trajectory tracking.

When the generalized dynamic and kinematic parameters are unknown, selecting $\boldsymbol\zeta$ in~(\ref{10}) as
\begin{equation} \label{31}
\boldsymbol\zeta = (\hat{\mathbf{J}}_m \hat{\mathbf{T}})^{+} [ -\hat{\mathbf{v}}_0 + \dot{\mathbf{x}}_{d} - \boldsymbol\Lambda_{x} \Delta \mathbf{x} - \hat{\mathbf{J}}_m \hat{\bar{\mathbf{H}}}^{+}_{bm} ( \hat{\mathbf{p}}_0 + \hat{\bar{\mathbf{H}}}_{b} \lambda_{b} \Delta\boldsymbol{\epsilon}_{bv} ) ]
\end{equation}
where $\hat{\mathbf{J}}_m$ is obtained by replacing the kinematic parameters in $\mathbf{J}_m$ with their estimates and $\hat{\mathbf{T}} = \mathbf{E}_{n \times n} - \hat{\bar{\mathbf{H}}}^{+}_{bm} \hat{\bar{\mathbf{H}}}_{bm}$, we obtain the following kinematic control law
\begin{equation} \label{24}
\dot{\boldsymbol\phi}^{\ast}_r = \hat{\mathbf{T}} \overbrace{(\hat{\mathbf{J}}_m \hat{\mathbf{T}})^{+} [ -\hat{\mathbf{v}}_0 + \dot{\mathbf{x}}_{d} - \boldsymbol\Lambda_{x} \Delta \mathbf{x} - \hat{\mathbf{J}}_m \hat{\bar{\mathbf{H}}}^{+}_{bm} ( \hat{\mathbf{p}}_0 + \hat{\bar{\mathbf{H}}}_{b} \lambda_{b} \Delta\boldsymbol{\epsilon}_{bv} ) ]}^{\boldsymbol\zeta} \\
+ \hat{\bar{\mathbf{H}}}^{+}_{bm} ( \hat{\mathbf{p}}_0 + \hat{\bar{\mathbf{H}}}_{b} \lambda_{b} \Delta\boldsymbol{\epsilon}_{bv} )
\end{equation}
where $\boldsymbol\Lambda_{x}$ is a constant symmetric positive definite matrix.

Premultiplying both sides of~(\ref{24}) by $\hat{\mathbf{J}}_{m}$, we get
\begin{equation} \label{25}
\hat{\mathbf{J}}_{m} \dot{\boldsymbol\phi}^{\ast}_r = \dot{\mathbf{x}}_{d} - \boldsymbol\Lambda_{x} \Delta \mathbf{x} - \hat{\mathbf{v}}_0.
\end{equation}

Combining~(\ref{4}) and~(\ref{25}), we have
\begin{equation} \label{26}
\dot{\mathbf{x}}_{d} - \boldsymbol\Lambda_{x} \Delta \mathbf{x} - \hat{\mathbf{J}}_{m} \dot{\boldsymbol\phi}^{\ast}_r - \hat{\mathbf{v}}_0
= \dot{\mathbf{x}} - \mathbf{J}_{b} \boldsymbol{\omega}_{b} - \mathbf{J}_{m} {\dot{\boldsymbol\phi}} - \mathbf{v}_0.
\end{equation}

Adding $\hat{\mathbf{J}}_{b} \boldsymbol{\omega}_{b} + \hat{\mathbf{J}}_{m} {\dot{\boldsymbol\phi}}$ to both sides of~(\ref{26}), and after some simple calculations, we obtain
\begin{equation} \label{27}
\hat{\mathbf{J}}_{b} \boldsymbol{\omega}_{b} + \hat{\mathbf{J}}_{m} (\dot{\boldsymbol{\phi}} - \dot{\boldsymbol\phi}^{\ast}_r)
- (\Delta\dot{\mathbf{x}} + \boldsymbol\Lambda_{x} \Delta \mathbf{x})
= \Delta\mathbf{J}_{b} \boldsymbol{\omega}_{b} + \Delta\mathbf{J}_{m} {\dot{\boldsymbol\phi}} + \Delta \mathbf{v}_0
\end{equation}
where $\Delta\mathbf{J}_{b} = \hat{\mathbf{J}}_{b} - \mathbf{J}_{b}$, $\Delta\mathbf{J}_{m} = \hat{\mathbf{J}}_{m} - \mathbf{J}_{m}$,
and $\Delta \mathbf{v}_0 = \hat{\mathbf{v}}_0 - \mathbf{v}_0$.

From~(\ref{5}), we can rewrite~(\ref{27}) as
\begin{equation} \label{28}
\underbrace{ \hat{\mathbf{J}}_{b} \boldsymbol{\omega}_{b} + \hat{\mathbf{J}}_{m} (\dot{\boldsymbol{\phi}} - \dot{\boldsymbol\phi}^{\ast}_r)
- (\Delta\dot{\mathbf{x}} + \boldsymbol\Lambda_{x} \Delta \mathbf{x}) }_{\mathbf{y}_{2}}
= \bar{\mathbf{Y}}_{k}(\boldsymbol\epsilon_{b}, \boldsymbol\phi, \boldsymbol\omega_{b}, \dot{\boldsymbol\phi}) \Delta\bar{\mathbf{a}}_{k}
\end{equation}
where $\Delta\bar{\mathbf{a}}_k = \hat{\bar{\mathbf{a}}}_k - \bar{\mathbf{a}}_k$ is the generalized kinematic parameter estimation error.
We assume that the position and the velocity of the end-effector are available from certain sensors.
Hence, $\mathbf{y}_{2}$ is measurable.
The generalized kinematic parameter estimate is updated by the gradient estimator of the standard form
\begin{equation} \label{29}
\dot{\hat{\bar{\mathbf{a}}}}_{k} = - \boldsymbol\Gamma_{k} \bar{\mathbf{Y}}^{\mathbf{T}}_{k} \mathbf{y}_{2}
\end{equation}
where $\boldsymbol\Gamma_{k}$ is a constant symmetric positive definite matrix.

Let us now introduce another sliding variable \cite{Slotine:1987}
\begin{equation} \label{30}
\mathbf{s}_{x} = \Delta\dot{\mathbf{x}} + \boldsymbol\Lambda_{x} \Delta \mathbf{x}.
\end{equation}

Based on the properties of the gradient estimator \cite{Slotine:1991}, the estimator~(\ref{16}) gives that $\hat{\bar{\mathbf{a}}}_{d} \in \mathcal{L}_{\infty}$ and $\mathbf{y}_{1} \in \mathcal{L}_{2}$, and the estimator~(\ref{29}) gives that $\hat{\bar{\mathbf{a}}}_{k} \in \mathcal{L}_{\infty}$ and $\mathbf{y}_{2} \in \mathcal{L}_{2}$.
According to the results in Section III-B, we know that $\boldsymbol\omega_b \in \mathcal{L}_2$, so the conclusion that $\mathbf{s}_{x} \in \mathcal{L}_2$ is reached from~(\ref{28}), which leads to the fact that $\Delta \mathbf{x} \in \mathcal{L}_{\infty}$ \cite[p.59]{Desoer:1975}.
Therefore, we obtain that $\boldsymbol{\zeta} \in \mathcal{L}_{\infty}$ from~(\ref{31}), which implies that $\dot{\boldsymbol\phi}^{\ast}_r \in \mathcal{L}_{\infty}$ if both $\hat{\mathbf{J}}_m \hat{\mathbf{T}}$ and $\hat{\bar{\mathbf{H}}}_{bm}$ have full row ranks.
Following the same procedure as that in Section III-B, we have that $\boldsymbol\omega_b \in \mathcal{L}_{\infty}$ and $\Delta\dot{\boldsymbol{\epsilon}}_{bv} \in \mathcal{L}_{\infty}$.
The boundedness of $\boldsymbol\omega_b$ and $\dot{\boldsymbol\phi}$ suggests that $\dot{\mathbf{x}} \in \mathcal{L}_{\infty}$ from~(\ref{5}) and $\mathbf{y}_{1} \in \mathcal{L}_{\infty}$ from~(\ref{14}).
The boundedness of $\dot{\mathbf{x}}_d$ gives that $\Delta\dot{\mathbf{x}} \in \mathcal{L}_{\infty}$, which yields that $\mathbf{y}_{2} \in \mathcal{L}_{\infty}$ from~(\ref{28}).
Consequently, we have that $\dot{\hat{\bar{\mathbf{a}}}}_{d} \in \mathcal{L}_{\infty}$ and $\dot{\hat{\bar{\mathbf{a}}}}_{k} \in \mathcal{L}_{\infty}$, which leads to the fact that $\dot{\boldsymbol\zeta} \in \mathcal{L}_{\infty}$ from~(\ref{31}).
Therefore, we obtain that $\ddot{\boldsymbol\phi}^{\ast}_r \in \mathcal{L}_{\infty}$ if both $\hat{\mathbf{J}}_m \hat{\mathbf{T}}$ and $\hat{\bar{\mathbf{H}}}_{bm}$ have full row ranks, which implies that $\ddot{\boldsymbol{\phi}} \in \mathcal{L}_{\infty}$ from the assumption that $\dot{\boldsymbol\phi} \equiv \dot{\boldsymbol\phi}_r^\ast$.
From the upper part of~(\ref{1}), we know that $\dot{\boldsymbol\omega}_{b} \in \mathcal{L}_{\infty}$, which implies that $\dot{\mathbf{y}}_{1} \in \mathcal{L}_{\infty}$ from~(\ref{21}),
and thus the signal $\mathbf{y}_{1}$ is uniformly continuous.
From the properties of square-integrable and uniformly continuous functions \cite[p.232]{Desoer:1975}, we have that $\mathbf{y}_{1} \to \mathbf{0}$ as $t \to \infty$.
Then, we obtain from~(\ref{15}) that $\mathbf{s}_{b} \to \mathbf{0}$ as $t \to \infty$
if $\hat{\mathbf{H}}_{b}$ is uniformly positive definite.
Thus, we have that
$\Delta\boldsymbol{\epsilon}_{bv} \to \mathbf{0}$ and $\Delta\boldsymbol\omega_b \to \mathbf{0}$ as $t \to \infty$ according to the result in \cite{Egeland:1994},
i.e., $\boldsymbol\omega_{b} \to \mathbf{0}$ and $\mathbf{R}_{b} \to \mathbf{R}_{bd}$ as $t \to \infty$.
So far, we have shown that the controller~(\ref{24}) with the estimators~(\ref{16}) and~(\ref{29}) can ensure the spacecraft attitude regulation.
Next, we will show that the controller~(\ref{24}) can also achieve trajectory tracking of the end-effector.

The boundedness of $\dot{\boldsymbol\omega}_b$ and $\ddot{\boldsymbol\phi}$ shows that $\ddot{\mathbf{x}} \in \mathcal{L}_{\infty}$ from the time derivative of the kinematics equation~(\ref{4}).
Thus, we have that $\Delta \ddot{\mathbf{x}} \in \mathcal{L}_{\infty}$ and then $\dot{\mathbf{y}}_{2} \in \mathcal{L}_\infty$, which means that $\mathbf{y}_{2}$ is uniformly continuous.
The fact that $\mathbf{y}_{2} \in \mathcal{L}_{2}$ and $\mathbf{y}_{2}$ is uniformly continuous leads to the result that
$\mathbf{y}_{2} \to \mathbf{0}$ as $t \to \infty$ \cite[p.232]{Desoer:1975}.
Since we have obtained that $\boldsymbol\omega_{b} \to \mathbf{0}$ as $t \to \infty$ and $\dot{\boldsymbol\phi} \equiv \dot{\boldsymbol\phi}_r^\ast$, the definition of $\mathbf{y}_{2}$ leads to the result that
$\mathbf{s}_{x} \to \mathbf{0}$ as $t \to \infty$.
According to the analysis in the work of \cite{Slotine:1987}, $\mathbf{s}_{x} \to \mathbf{0}$ as $t \to \infty$
implies that $\Delta \mathbf{x} \to \mathbf{0}$ and $\Delta\dot{\mathbf{x}} \to \mathbf{0}$ as $t \to \infty$,
which means that $\mathbf{x} \to \mathbf{x}_{d}$ and $\dot{\mathbf{x}} \to \dot{\mathbf{x}}_{d}$ as $t \to \infty$.

We summarize the above analysis as the following theorem.

\begin{theorem}
The kinematic control law~(\ref{24}) and the parameter adaptation laws~(\ref{16}), (\ref{29}) achieve the regulation of the spacecraft attitude and the convergence of the FFSM end-effector tracking errors provided that there exists a fast enough dynamic control law so that $\dot{\boldsymbol{\phi}} \equiv \dot{\boldsymbol\phi}^{\ast}_r$.
That is, $\boldsymbol{\omega}_{b} \to \mathbf{0}$, $\mathbf{R}_{b} \to \mathbf{R}_{bd}$, $\Delta \mathbf{x} \to \mathbf{0}$ and $\Delta\dot{\mathbf{x}} \to \mathbf{0}$ as $t \to \infty$.
\end{theorem}

\section{SIMULATION RESULTS}

\begin{figure}
\centering
\includegraphics[width=8.4cm]{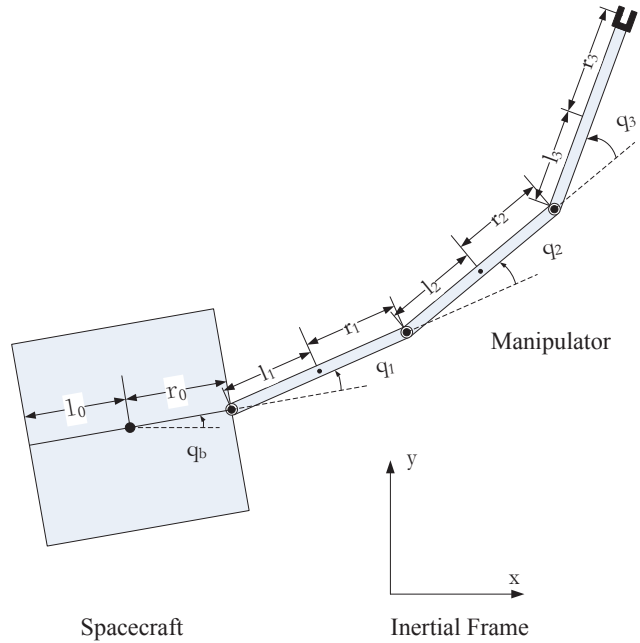}    
\caption{Three-DOF planar free-floating space manipulator.}
\label{fig:space manipulator}
\end{figure}

In this section we present simulation results for the proposed adaptive control law via a typical three-DOF planar space manipulator (Fig.~\ref{fig:space manipulator}).
For the system considered here, the number of manipulator joints $n=3$, and the number of task variables for the spacecraft task is $d_{1} = 1$.
Since the end-effector moves in the plane, the number of task variables for the end-effector task is $d_{2} = 2$. Thus, $n=d_{1}+d_{2}$, which implies that the DOFs of the manipulator are just enough for realizing the spacecraft attitude regulation and manipulator end-effector tracking.

The physical parameters of the space manipulator are listed in Table~\ref{tb:Free_Floating_Space_Manipulator parameters}, where $m_{i}$ is the mass of the $i$-th rigid body, $I_{i}$ is the moment of inertia of the $i$-th body about the center of mass (CM), $l_{i}$ and $r_{i}$ are shown as Fig.~\ref{fig:space manipulator}, $i=0,1,2,3$, and the 0-th body denotes the spacecraft.
The expressions for $\bar{\mathbf{H}}_{b}$, $\bar{\mathbf{H}}_{bm}$, $\mathbf{J}_{b}$, $\mathbf{J}_{m}$ can be found in \cite{Nenchev:1988}, and the kinematic and dynamic parameters for estimation are chosen based on the kinematic and dynamic models given in \cite{Nenchev:1988}. The sampling period used in the following simulations is set as 2 ms.

\begin{table}[hb]
\begin{center}
\caption{Physical parameters of the space manipulator} \label{tb:Free_Floating_Space_Manipulator parameters}
\begin{tabular}{ccccc}
$i$-th body & $m_i(\mathrm{kg})$ & $I_i(\mathrm{kg \cdot m^2})$ & $l_i(\mathrm{m})$ & $r_i(\mathrm{m})$ \\\hline
0 & 61.2 & 26.1120 & 0.80 & 0.80 \\
1 & 6.3  & 1.0290 & 0.70 & 0.70 \\
2 & 5.4  & 0.8820 & 0.70 & 0.70 \\
3 & 5.1  & 0.8330 & 0.70 & 0.70 \\ \hline
\end{tabular}
\end{center}
\end{table}

The desired end-effector trajectory of the FFSM is given by
\begin{equation}
\mathbf{x}_d = {\begin{bmatrix} 3.7+0.3\cos(\pi t) \\ 0.2+0.3\sin(\pi t) \end{bmatrix}}.  \nonumber
\end{equation}
The desired value of the spacecraft attitude is set as zero, i.e., $\mathbf{q}_{bd}=0$. The initial values of the position of the CM of the spacecraft, the spacecraft attitude, the manipulator joint position, and the FFSM end-effector position are set as $\mathbf{R}_{C0}(0) = {\begin{bmatrix} 0~~0 \end{bmatrix}}^{\mathbf{T}}$, $\mathbf{q}_b(0) = 0$, and $\mathbf{q}_m(0) = {\begin{bmatrix} \pi/3 ~~ -2\pi/3 ~~ \pi/3 \end{bmatrix}}^{\mathbf{T}}$, respectively.
The initial velocities are determined as $\dot{\mathbf{R}}_{C0}(0) = {\begin{bmatrix} 0.1~~0.1 \end{bmatrix}}^{\mathbf{T}}$, $\dot{\mathbf{q}}_b(0) = -0.05$,
and $\dot{\mathbf{q}}_m(0) = {\begin{bmatrix} 0.05 ~~ -0.01 ~~ 0.09 \end{bmatrix}}^{\mathbf{T}}$.
The initial values of the generalized kinematic and dynamic parameter estimates are chosen as
\begin{equation}
\hat{\mathbf{a}}_{k}(0) = {\begin{bmatrix} 2 ~~ 3 ~~ 3 ~~ 3 ~~ 0 ~~ 0 \end{bmatrix}}^{\mathbf{T}},  \nonumber
\end{equation}
\begin{equation}
\hat{\mathbf{a}}_{d}(0) = {\begin{bmatrix} 30 ~~ 20 ~~ 3 ~~ 3 ~~ 3 ~~ 5 ~~ 100 ~~ 60 ~~ 30 ~~ 2 ~~ 0 \end{bmatrix}}^{\mathbf{T}}.  \nonumber
\end{equation}
The actual values of the kinematic parameter and dynamic parameter are calculated using the physical parameters given
in Table~\ref{tb:Free_Floating_Space_Manipulator parameters}, i.e.,
\begin{equation}
\mathbf{a}_k = {\begin{bmatrix} 0.6277 ~~ 1.1550 ~~ 1.2600 ~~ 1.3542 \end{bmatrix}}^{\mathbf{T}},  \nonumber
\end{equation}
\begin{equation}
\begin{aligned}
\mathbf{a}_d = [ 11.9952 ~~ \allowbreak 12.6126 ~~ \allowbreak 4.1234 ~~ \allowbreak 4.4982 ~~ \allowbreak 6.8544 \\ ~~ \allowbreak 2.2409
~~ \allowbreak 69.7260 ~~ \allowbreak 35.1779 ~~ \allowbreak 15.1638 ~~ \allowbreak 3.1686]^{\mathbf{T}}.  \nonumber
\end{aligned}
\end{equation}
The actual value of the initial angular momentum is $\mathbf{p}_0 = -1.6467$,
and the actual value of $\mathbf{v}_0$ is $\mathbf{v}_0 = {\begin{bmatrix} 0.0988 ~~ 0.0943 \end{bmatrix}}^{\mathbf{T}}$.
The controller parameters are determined as
$\lambda_{b} = 60$, $\boldsymbol\Lambda_{x} = 20\mathbf{E}_{2 \times 2}$,
$\boldsymbol\Gamma_d = \mathrm{diag}([30 ~~ \allowbreak 30 ~~ \allowbreak 10 ~~ \allowbreak 10 ~~ \allowbreak 10 ~~ \allowbreak 10 ~~ \allowbreak 10 ~~ \allowbreak 10 ~~ \allowbreak 10 ~~ \allowbreak 10 ~~ \allowbreak 1])$,
and $\boldsymbol\Gamma_k = \mathrm{diag}([20 ~~ 20 ~~ 20 ~~ 20 ~~ 2 ~~ 2])$.

\begin{figure}
\centering
\includegraphics[width=8.4cm]{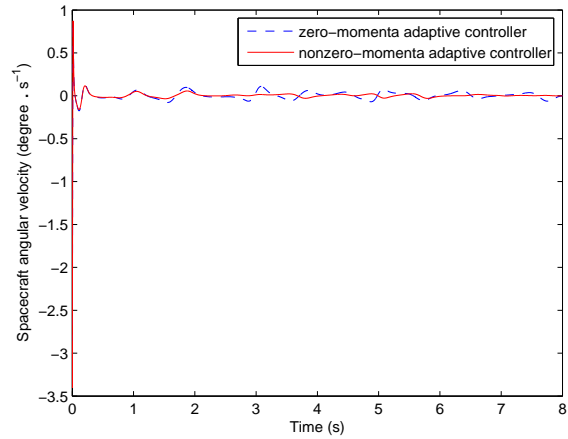}    
\caption{Spacecraft angular velocity under the zero-momenta and nonzero-momenta adaptive controller.}
\label{fig:angular velocity of the base_COMPARE}
\end{figure}

\begin{figure}
\centering
\includegraphics[width=8.4cm]{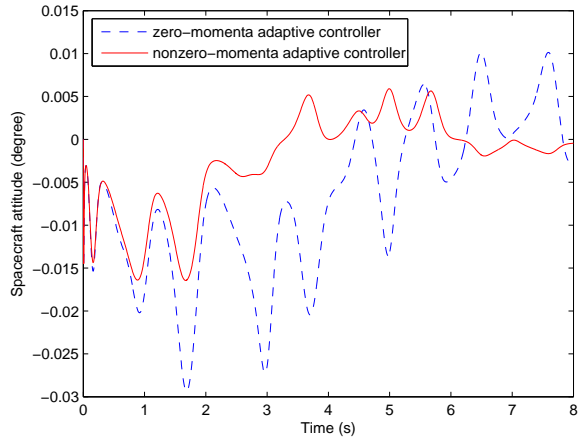}    
\caption{Spacecraft attitude under the zero-momenta and
nonzero-momenta adaptive controller.}
\label{fig:base attitude_COMPARE}
\end{figure}

\begin{figure}
\centering
\includegraphics[width=8.4cm]{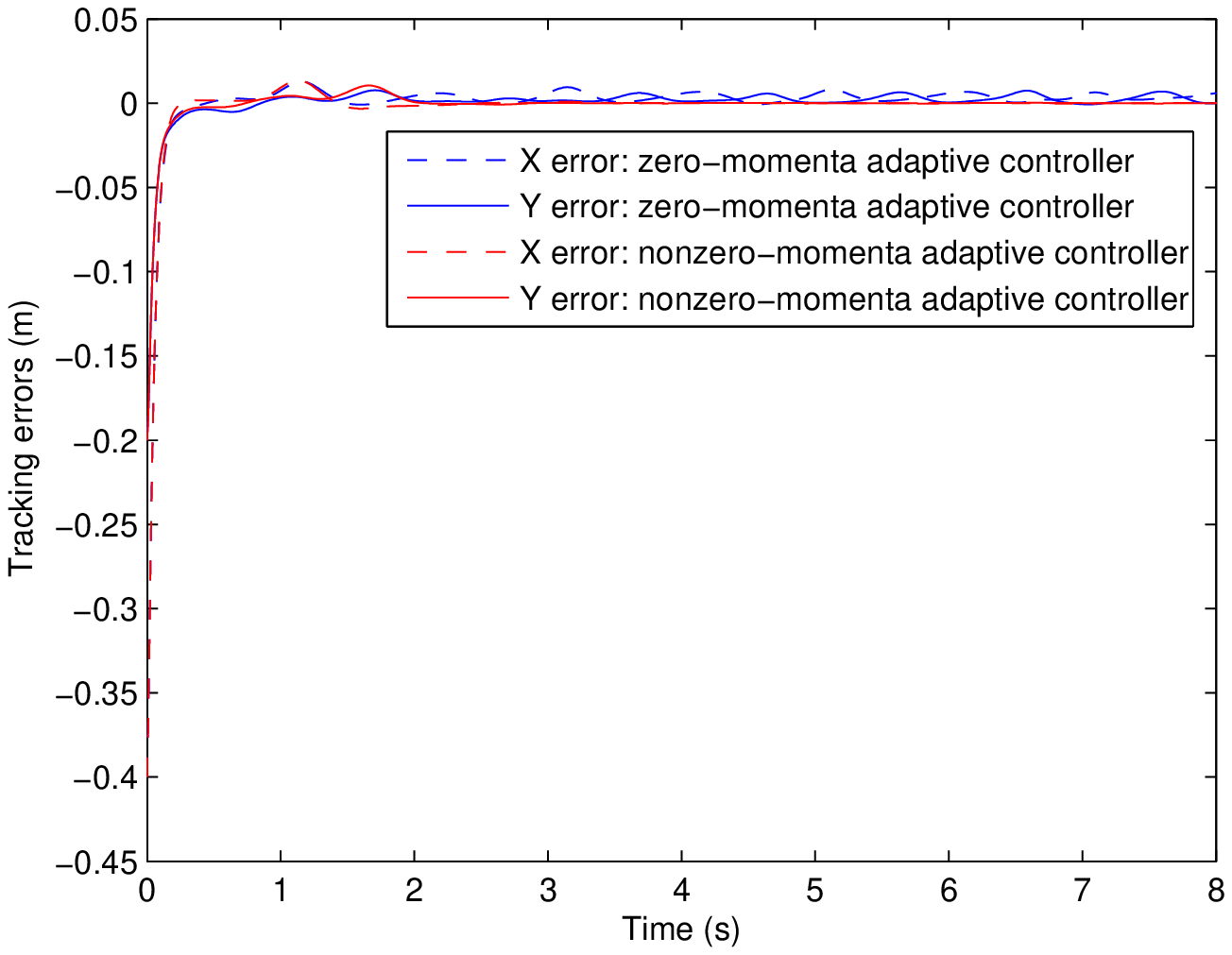}    
\caption{FFSM end-effector tracking errors under the zero-momenta and nonzero-momenta adaptive controller.}
\label{fig:tracking error_COMPARE}
\end{figure}

We call the controller proposed in our preliminary work~\cite{Xu:2013}, which is the case that the effects of the nonzero initial linear and angular momenta are not considered in the controller~(\ref{24}), the ``zero-momenta adaptive controller".
As its counterpart, the controller~(\ref{24}) in this paper
is called the ``nonzero-momenta adaptive controller".
In order to compare the performance of these two controllers, the responses under both the controllers are shown in Figs.~\ref{fig:angular velocity of the base_COMPARE}-\ref{fig:tracking error_COMPARE}.

Figs.~\ref{fig:angular velocity of the base_COMPARE}-\ref{fig:tracking error_COMPARE} give the angular velocity and the attitude of the spacecraft, and the FFSM end-effector tracking errors under these two controllers.
Fig.~\ref{fig:angular velocity of the base_COMPARE} shows that the angular velocity of the spacecraft tends to zero under these two controllers, and no significant differences are noticed.
From Fig.~\ref{fig:base attitude_COMPARE}, we see that the spacecraft attitude tends to zero (i.e., the desired value), and that the response under the ``nonzero-momenta adaptive controller" is smoother than that of its counterpart.
The greater difference lies in the tracking
errors of the end-effector, as is shown in Fig.~\ref{fig:tracking error_COMPARE}, and the large tracking errors under the zero-momenta adaptive controller is caused by the fact that it does not take into account the nonzero momenta.
The comparison between them illustrates the effectiveness of the ``nonzero-momenta adaptive controller" proposed in this paper.

\begin{figure}
\centering
\includegraphics[width=8.4cm]{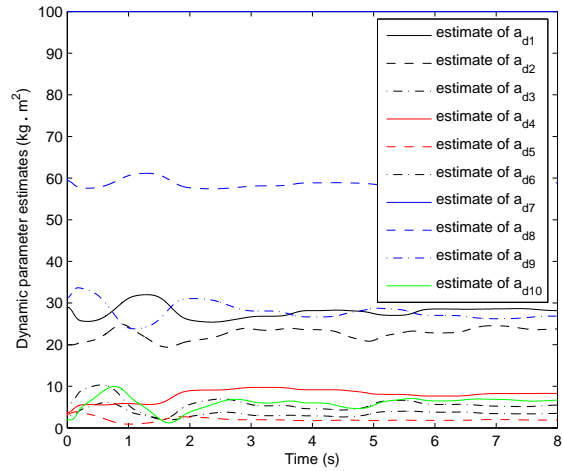}    
\caption{Dynamic parameter estimates.}
\label{fig:Dyn_para_estimation}
\end{figure}

\begin{figure}
\centering
\includegraphics[width=8.4cm]{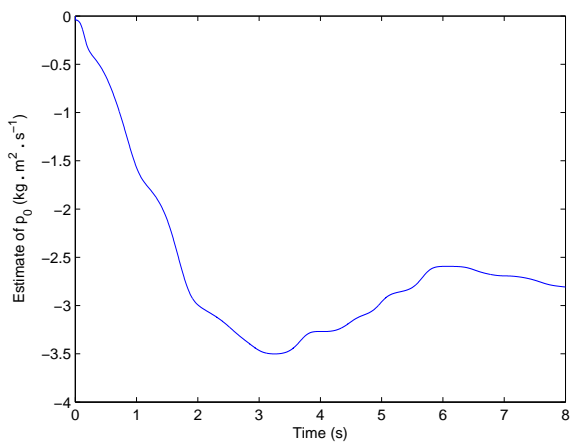}    
\caption{Estimate of the initial angular momentum of the FFSM (i.e., $\mathbf{p}_0$).}
\label{fig:Estimation_L0}
\end{figure}

\begin{figure}
\centering
\includegraphics[width=8.4cm]{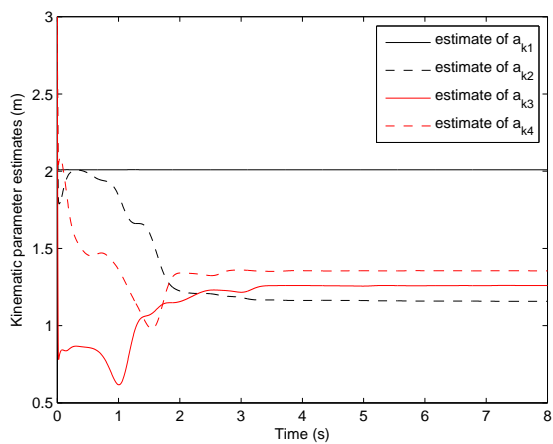}    
\caption{Kinematic parameter estimates.}
\label{fig:Kin_para_estimation}
\end{figure}

\begin{figure}
\centering
\includegraphics[width=8.4cm]{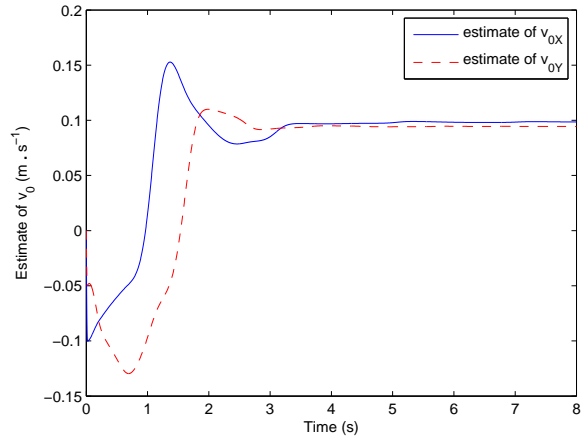}    
\caption{Estimate of the initial velocity of the CM of the FFSM (i.e., $\mathbf{v}_0$).}
\label{fig:Estimation_ll0}
\end{figure}

\begin{figure}[h]
\centering
\includegraphics[width=8.4cm]{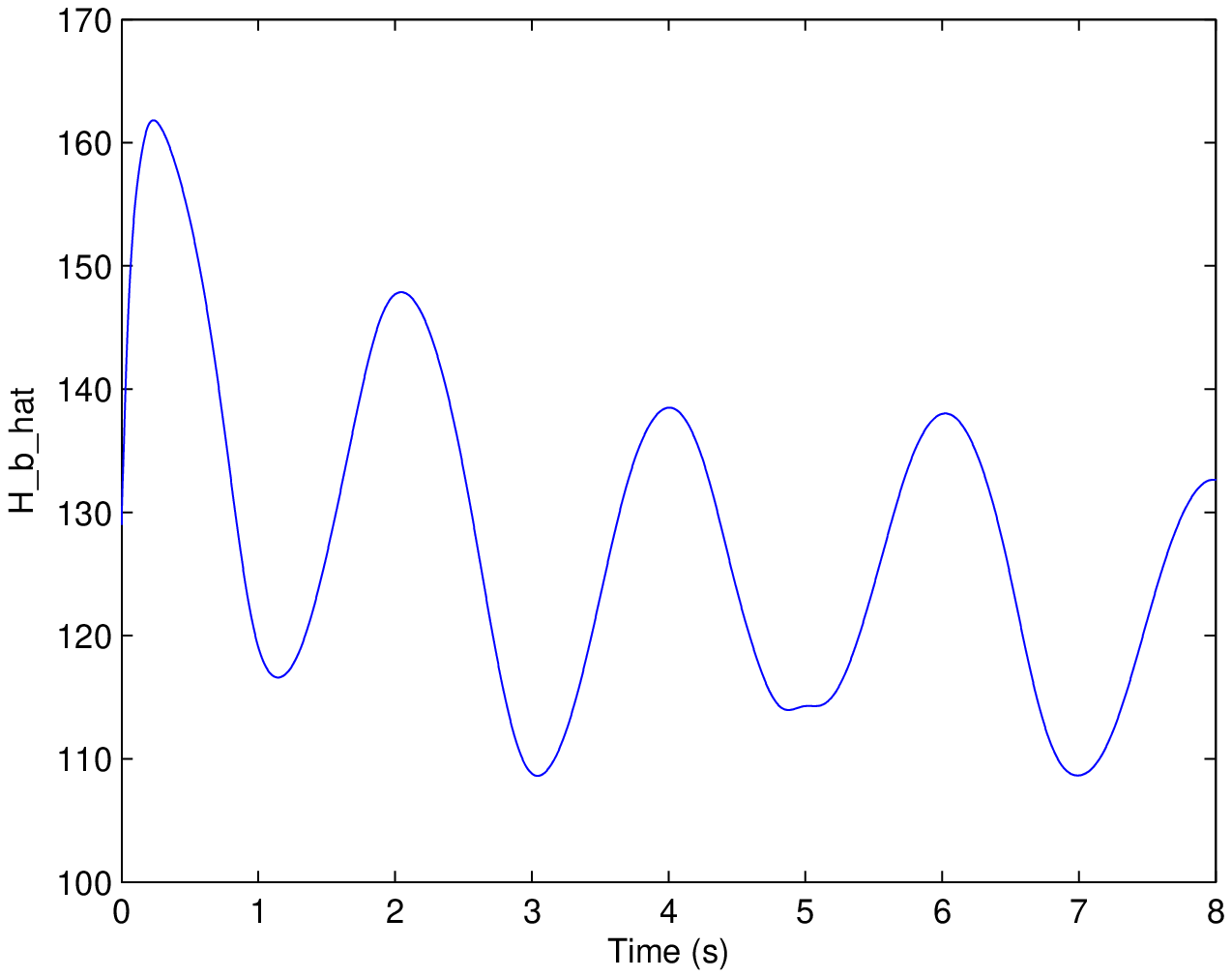}    
\caption{Estimate of $\mathbf{H}_b$.}
\label{fig:estimation_H_b}
\end{figure}

Under the proposed controller~(\ref{24}), we estimate both the generalized dynamic parameters $\bar{\mathbf{a}}_d$ and the generalized kinematic parameters $\bar{\mathbf{a}}_k$, taking into consideration the nonzero initial momenta.
The parameter estimates under the proposed adaptive controller are shown in Figs.~\ref{fig:Dyn_para_estimation}-\ref{fig:estimation_H_b}.

Since the magnitude of the estimate of the initial angular momentum is rather smaller compared with those of the dynamic parameter estimates, they are plotted in separate figures (i.e., Fig.~\ref{fig:Dyn_para_estimation} and Fig.~\ref{fig:Estimation_L0}).
For the same reason, the estimate of the initial velocity of the CM of the FFSM and those of the kinematic parameters are separately shown in Fig.~\ref{fig:Kin_para_estimation} and Fig.~\ref{fig:Estimation_ll0}.
Fig.~\ref{fig:estimation_H_b} presents the estimate of the inertia matrix $\mathbf{H}_b$.
It shows that the estimated spacecraft inertia is always positive definite (here, $\hat{\mathbf{H}}_b$ is a $1 \times 1$ matrix), so the parameter projection algorithm is not required here.

\section{Conclusion}

In this work, an adaptive zero reaction joint motion controller at velocity level has been presented for free-floating space manipulators with uncertain kinematics and dynamics. The RNS based kinematic control law can not be linearly parameterized, which is a great challenge for developing an adaptive controller. Giving insight into the structure of the RNS based controller, we have skillfully developed a linear expression which facilitates the deriving of the adaptive controller.
By exploiting the feature of the vector $\boldsymbol\zeta$, we propose an adaptive controller that can guarantee both the end-effector trajectory tracking and spacecraft attitude regulation.
In our future work, we will investigate the adaptive zero reaction controller design at the acceleration level.

\section*{Acknowledgment}

The authors would like to thank Prof. Chunling Wei, Dr. Yong Wang and Dr. Xindong Li for their suggestions on improving the quality of this paper.

\ifCLASSOPTIONcaptionsoff
  \newpage
\fi

\end{document}